# Functional annotation of creeping bentgrass protein sequences based on convolutional neural network

Han-Yu Jiang[a,b], Jun He[a*]

[a] School of Physics and Technology, Nanjing Normal University, Nanjing, Jiangsu 210097, China

[b] Sino-U.S. Center for Grazingland Ecosystem Sustainability/Pratacultural Engineering Laboratory of Gansu Province/ Key Laboratory of Grassland Ecosystem, Ministry of Education/College of Pratacultural Science, Gansu Agricultural University, Lanzhou, Gansu 730070, China

## Abstract

**Background:**

Creeping bentgrass (Agrostis soionifera) is a perennial grass of Gramineae, belonging to cold season turfgrass, but has poor disease resistance. Up to now, little is known about the induced systemic resistance (ISR) mechanism, especially the relevant functional proteins, which is important to disease resistance of turfgrass. Achieving more information of proteins of infected creeping bentgrass is helpful to understand the ISR mechanism.

**Results:**

With BDO treatment, creeping bentgrass seedlings were grown, and the ISR response was induced by infecting *Rhizoctonia solani*. High-quality protein sequences of creeping bentgrass seedlings were obtained. Some of protein sequences were functionally annotated according to the database alignment while a large part of the obtained protein sequences was left non-annotated. To treat the non-annotated sequences, a prediction model based on convolutional neural network was established with the dataset from Uniport database in three domains to acquire good performance, especially the higher false positive control rate. With established model, the non-annotated protein sequences of creeping bentgrass were analyzed to annotate proteins relevant to disease-resistance response and signal transduction.

**Conclusions:**

The prediction model based on convolutional neural network was successfully applied to select good candidates of the proteins with functions relevant to the ISR mechanism from the protein sequences which cannot be annotated by database alignment. The waste of sequence data can be avoided, and research time and labor will be saved in further research of protein of creeping bentgrass by molecular biology technology. It also provides reference for other sequence analysis of turfgrass disease-resistance research.

---

[*] Corresponding author, email: junhe@njnu.edu.cn

# 1. Introduction

Creeping bentgrass (*Agrostis soionifera*) is a perennial grass of Gramineae, belonging to cold season turfgrass. Due to its excellent characteristics, such as, strong adaptability, good ornamental, it is a preferred grass species in golf course, lawn tennis court, courtyard, park and other green areas. However, creeping bentgrass with shallow adventitious roots has poor disease-resistance. For example, it is susceptible to coin spot and brown spot. The innate immunity can be induced in plant, which relies on a surprisingly complex response mechanism to recognize and counteract different invaders. The induced physical and chemical barriers are activated to effectively combat invasion by microbial pathogens, as well as inducible defensive mechanisms upon attack [1,2]. Among them, the induced systemic resistance (ISR) is often activated by plant growth promoting bacteria in soil rhizosphere, and has broad-spectrum resistance to bacteria, fungi and pathogens [3,4].

Since without disease resistance-inducing factor the resistance of plants may not be induced [5], Butanediol (BDO) is often adopted as a new type of disease resistance-inducing factor, which provides durable disease resistance. ISR produced by BDO effectively inhibits grass leaf diseases [6,7]. Studies have shown that many resistance proteins enter the nucleus to activate the immune response and triggers the signal transduction pathway, including resistance signal activation, transcription factor regulation and hormone signal pathway activation [8]. For instance, a number of preliminary proteome analyses in rice successfully identified some known pathogenesis-related proteins that accumulate abundantly after Jasmonic acid treatment or inoculation by the pathogenic fungus *M. grisea* [9,10]. Oh et al. [11] analyzed the secreted protein encoding the lipase with antimicrobial activity in Arabidopsis. However, except for these few preliminary studies, the study about the proteins relevant to the ISR mechanism is still scarce, especially for the turfgrass. One major reason is that many proteins identified and analyzed involving in signaling processes are below the threshold of detection [12]. Hence, more efforts are worth and urgent to be put into the study of disease-resistance related proteins.

In our previous work [13], BDO was used to induce ISR resistance in creeping bentgrass infected with *Rhizoctonia solani*, and a genetic research of creeping bentgrass by the transcriptome analysis was performed to analyze ethylene-dependent signal transduction pathways involved in ISR mechanisms. In that work, only the sequences annotated by the database alignment were analyzed. However, there are a large number of protein sequences, which were not aligned in seven databases, or aligned but not annotated. Since these sequences were non-annotated, analysis cannot be performed, so not reported in previous work. Considering the important role played by the protein in ISR mechanism, in this work, we will provide an explicit analysis about these protein sequences of creeping bentgrass. It is very helpful for fully understanding of the ISR mechanism and further study of the disease resistance of creeping bentgrass if we can annotate these protein sequences correctly even not exhaustively.

The function of protein is usually analyzed and annotated by biochemical experiments, which are time- and labor-consuming. At present, the number of protein sequences in UniPort database exceeds 100 million, and still increases rapidly [14]. The traditional methods are not enough to make up the increasing gap between the requirement and the speed of protein annotation by experimental means [15]. Therefore, the protein function prediction methods, such as machine learning, were proposed and widely adopted in the research [16]. However, traditional computational methods have disadvantages such as high false positive control rate and low accuracy. In the recent year, the deep learning technology becomes an important method in the field of protein research [17,18] while it is scarcely applied to the study of grass. In Ref. [19], an accurate and stable function prediction model was built to extract protein features with the convolutional neural network (CNN). It may solve the shortcomings of other methods, especially false positive control rate.

In the current work, the protein sequences from the creeping bentgrass seedlings with BDO treatment will be reported. To provide more helpful information about the protein sequences obtained, with the deep learning algorithm, we performed following analysis about non-annotate protein sequences of creeping bentgrass, (1) Based on CNN and protein binary encoding representation strategy, a functional prediction model was established following Ref. [19] with some modifications and adjusted to achieve high false positive control rate with annotated protein sequences from some Gene Ontology (GO) terms, which were collected from the Uniport database; (2) The established model was applied to non-annotated protein sequences of creeping brntgrass to predict functional classification; (3) The prediction model was further applied to select the proteins relevant to disease-resistance and signal transduction from non-annotated protein sequences of creeping bentgrass. Such treatment avoids waste of data, and supplements the analysis of proteins of creeping bentgrass. The research lays foundation for further mining ISR response proteins of creeping bentgrass and exploring the disease-resistance mechanism of turfgrass.

## 2. Materials and methods

### 2.1 Plant growth conditions and production of sequencing libraries

Seeds from creeping bentgrass 'PennA-4' (Chinese Academy of Agricultural Sciences) were grown by modified method of Kroes et al [13,20]. The surface of seeds was disinfected with 70% ethanol for 1 min, disinfected with 15% sodium hypochlorite for 5min, cleaned with sterile water for 10 min, and finally dried with filter paper. Seeds were sown in 50-mL culture flask with 10 mL of MS medium containing 100 µmol·L$^{-1}$ of BDO. About 20 seedlings per flask were cultured in a growth chamber at 22 °C under 100 µEm$^{-2}$s$^{-1}$ light. The experimental materials were seedlings cultured for twelve-day-old under the above conditions. *Rhizoctonia solani* (#3.2888 from China General Microbiological Culture Collection Center) in PDO liquid medium (potato 200 g·L$^{-1}$, glucose 20 g·L$^{-1}$) was shaking culture for 2-3 day with 120 r·min$^{-1}$ at 25°C. Concentration of the bacterial sample was a final OD$_{340}$ of 0.8. Roots of seedlings were directly sprayed with 2 mL of the bacterial fluid. The brown blotch symptoms of creeping bentgrass seedlings were observed, and the mycelium began to grow after 3–5-day post-inoculation. After 24, 48 and 72 h post-inoculation, the seedlings with different treatment were removed, and the leaves were cut. The treated materials were tested and analyzed, and all the samples were mixed and spliced (Eukaryotic Non-reference Transcriptome). The transcriptome analysis was performed by Illumina Sequencing. Sequencing libraries were produced by NEBNext UltraTM RNA Library Prep Kit (NEB, San Diego, CA, USA), and each sample attributes sequences for index codes.

## 2.2 Establishing of protein function prediction model

### 2.2.1 Constructing the data sets of training and testing

At first, we should construct database for establishing the prediction model under CNN frame. Some of obtained protein sequences of creeping bentgrass can be aligned and classified into the GO terms. Excluding some GO terms with too few protein sequences, we chose 7 terms in cellular component (CC) domain, 10 terms in molecular function (MF) domain, and 12 terms in biological process (BP) domain. The annotated protein sequences with these GO terms were collected from the UniPort database (see Fig. 1).

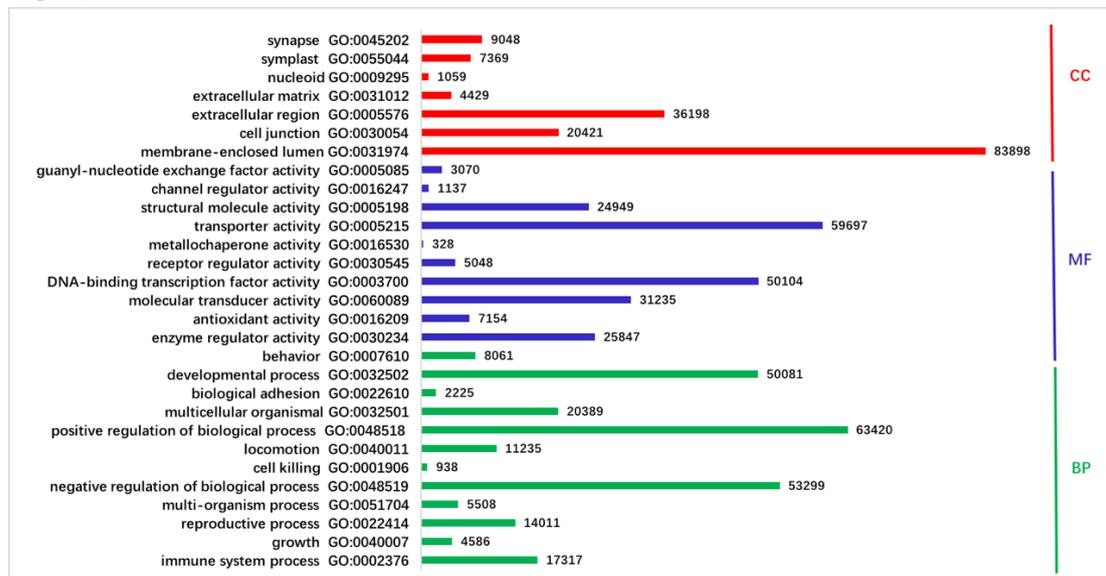

Fig. 1. The number of protein sequences in every GO term used in training.

With the protein sequences collected, we constructed positive and negative data for establishing prediction model by adopting binary classification, which would be also used to treat protein sequences of creeping bentgrass. For a GO term studied, we considered the proteins in this GO term as positive data. The negative data were selected from the left GO terms after removing the repeated sequences. To avoid overemphasizing one of all left GO terms, for a GO term studied with $N$ sequences, we selected the sequences from the left GO terms in order until 3N sequences were selected. A data set with $4N$ sequences was obtained. After shuffling, 60% of $4N$ proteins were used as training set, 20% as testing set, and 20% as valuation set. Here, imbalance binary classification was adopted to emphasize the negative data. Such treatment can improve the false positive control rate that we focused on in the current work, but lower sensitivity, which is less important in the present work.

**2.2.2 Prediction model based on CNN**

In the current work, we adopted a deep learning algorithm, CNN, to analyze the protein sequences. The protein is expressed as a one-dimensional sequence of amino acids, which is quite analogous to the sentence classification. We chose an explicit CNN frame, textCNN [21], which has been successfully applied to analyze the text, and to study the proteins [19], which we followed in the current work. The model was implemented with the Tensorflow3 library and the python programming language with some modifications to get best performance for the data sets considered in the current work. The binary cross-entropy loss function was adopted in all models training, and the Adam [22] optimizer with default parameters was used for the optimization during back-propagation. The weight parameters were initialized with the He initialization method [23], and biases were initialized to zero.

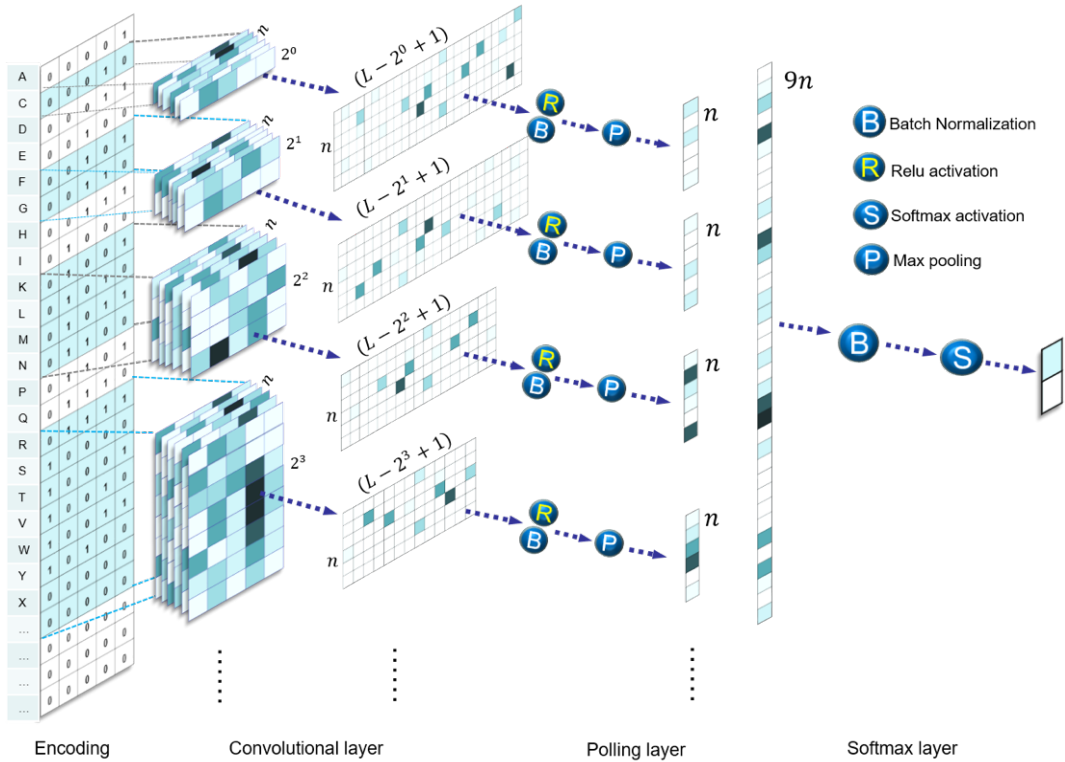

Fig. 2. The workflow of CNN adopted in the current work.

For a protein sequence, we need first encode the amino acids into binary vectors. Because only about twenty amino acids were discovered, we encoded an amino acid to a 5-bit binary vector as shown in Fig. 2. For example, the alanine was encoded as [0,0,0,0,1]. In the current work we did not distinguish the rare and undetermined amino acids, and encoded them all as [0,0,0,0,0]. The lengths of protein sequences are different while the CNN requires a fixed length. We considered the proteins of sequence length less than $L = 800$ amino acids, which constitute the majority (>95%) of the protein sequences in the studied GO terms (and almost 100% of non-annotated sequences of creeping bentgrass). For the protein sequences less than 800 amino acids, the left positions were complemented by binary vector [0,0,0,0,0]. With such encoding, a protein sequence was converted into an $L \times 5$ matrix.

After encoding, the convolution layer with He normalization as kernel initializer was adopted to extract the information from the digitized protein sequences as shown in Fig. 2. To obtain the information on different length levels, convolution kernels with different sizes were adopted with layers.Conv2D function in Tensorflow3. In the current work, we chose $n = 120$ convolution kernels with size as $2^k \times 5$ with $k = 0, 1, 2, \ldots, 8$. After convolution, nine $(L - 2^k + 1) \times n$ arrays were obtained as

$$a_{ij}^k = \sum_{m=1}^{2^k} \sum_{l=1}^{5} \left( X_{(m+i-1)l} * W_{ml}^j \right) + b_i^j,$$

where $j$ is for the number of kernels with size as $2^k \times 5$. To accelerate the learning speed, the batch normalization was applied with parameters momentum and epsilon being 0.99 and 0.001, respectively, which was followed by the ReLu activation. The max pooling was adopted by selecting the maximum in $(L - 2^k + 1)$ elements of $a_{ij}^k$ for certain $k$ and $j$. Then, the $k$ vectors with the same size $n$ were concatenated to a vector with size $kn = 1080$. Additional fully connected layer was not applied here because it was found not helpful to improve the results and made the model hard to converge. Instead, modified from Ref. [19], after batch normalization, we adopted the layers.dense function to provide the classification probability. It includes a layer with 1080 input neurons and 2 output neurons. The softmax activation was then adopted to transfer the values of the two output neurons $a_{1,2}$ to $y_{1,2}$ with differentiable softmax function

$$y_{1,2} = \frac{e^{a_{1,2}}}{e^{a_1} + e^{a_2}}.$$

One can find that $y_{1,2}$ is value between 0 and 1 and $y_1 + y_2 = 1$, which is just the classification probability. If the possibility $y_1 > 0.5$, the sequence will be classified into the GO term. Here the L2 regularization was also adopted to avoid overfitting.

## 3. Results and discussions

### 3.1 Model's performance with the protein sequences from the Uniport database

When establishing the prediction model above, we trained the model by the protein sequences collected from the Uniport database in 29 GO terms, which numbers were shown in Fig. 1. The model and parameters were adjusted to obtain the best performance of false positive control rate because in the current work we want to establish a prediction model to select proteins with certain function correctly but not exhaustively. To evaluate the performance of the model, we introduce five widely-used measurements, sensitivity (SE), specificity (SP), precision (PR), accuracy (AC) and Matthews correlation coefficient (MCC), which explicit definitions are given in Additional file 1. Amongst five measurements, the specificity (SP) is the most important factor because it reflects the false positive. In Fig. 3, we presented the violin plots to show the overall picture of the results for three domains.

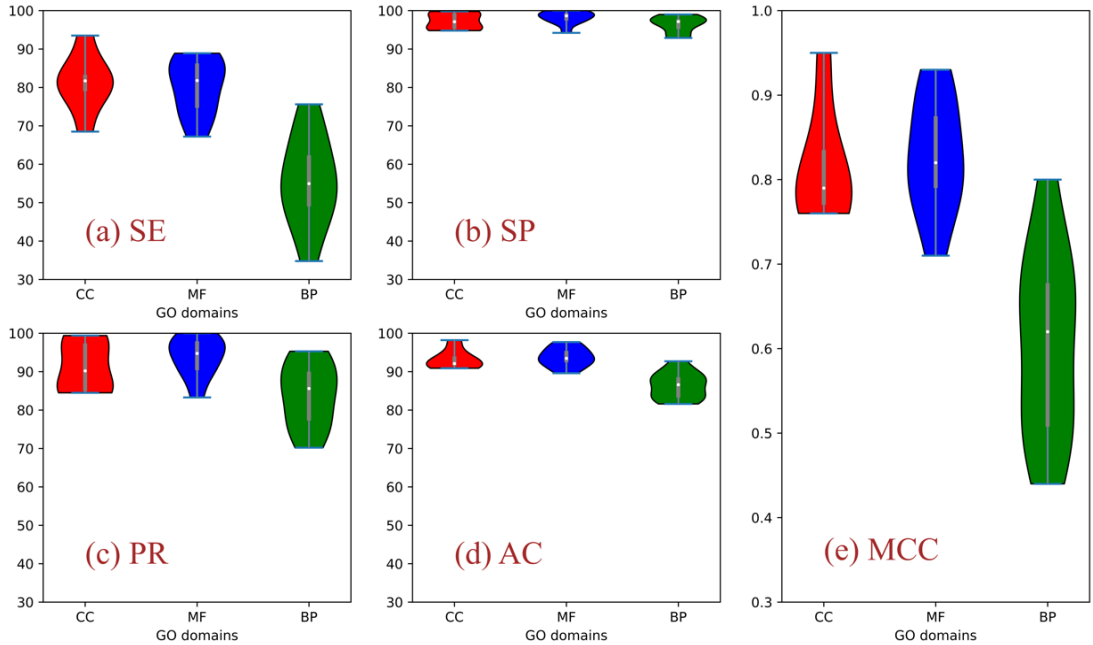

Fig. 3. Violin plots for the performance of prediction model in three domains.

Generally speaking, the model works well for the GO terms in three domains, especially for the GO terms in MF domain, which is comparable to the model in Ref. [19] where the studied GO terms are also in MF domain. Amongst five measurements, SP values are quite well for three domains (Fig. 3b), which satisfies our requirement to ensure the correctness of functional annotation of protein. Besides, smaller MCC values for BP domain than other two domains suggests that the functions of proteins in BP domain are more complex and harder to be learned by the prediction model.

**3.2 Screening of non-annotated sequences**

To annotate the protein sequences of the transcripts obtained in Section 2.1, based on four databases, NR, KOG, SwissProt, and KEGG, the BLASTALL package (release 2.2.28) [24] from the NCBI was adopted with a significant threshold of E-value $10^{-5}$. The KOBAS software in the KEGG pathways was used to check the statistical enrichment of differential expression genes (DEGs) [25]. The Blast2go v2.5 software (Biobam, Spain) was used to functionally categorize the sequences based on Gene Ontology (GO) with an E-value filter $1 \times 10^{-6}$. The total number of annotated protein sequences is 208,672. GO classification associations, combined with statistically transcripts were uploaded on the National Center for Biotechnology Information (NCBI) Sequence Read Archive (SRA) under accession number SRR5658390. The annotated protein sequences can be found in these data. In the followings, we will focus on the protein sequences which is not annotated.

The non-annotated sequences used in our study referred to the sequences that were not aligned with databases, or the sequences that were aligned but no predicted results. According to the database alignment, 118,856 amino acid sequences were not annotated. All non-annotated amino acid sequences are provided in the Additional file 2. The exploration of this part of protein functions has great significance for further understanding the ISR mechanism of creeping bentgrass. The annotation results of these non-annotated protein sequences with established model are given in the followings.

## 3.3 Functional annotation of non-annotated sequences of creeping bentgrass by GO terms in three domains

In the above, we established the prediction model with good performance, especially for the false positive control rate, by training the GO terms in three domains. With the trained model, we can analyze non-annotated protein sequences of creeping bentgrass and classify them into the GO terms considered above. The protein sequences with the function of certain GO term selected by prediction model are provided in the Additional file 3. In Table 1, the number of the proteins belonging to a GO term and the ratio of the number to the total number of the protein sequences are listed.

**Table 1**
The number and ratio of the proteins belonging to a GO term predicted from non-annotated protein sequences of creeping bentgrass.

| CC | | | MF | | | BP | | |
|---|---|---|---|---|---|---|---|---|
| GO ID | Num | % | GO ID | Num | % | GO ID | Num | % |
| GO:0045202 | 10095 | 8.49 | GO:0005085 | 1026 | 0.86 | GO:0007610 | 1410 | 1.19 |
| GO:0055044 | 1177 | 0.99 | GO:0016247 | 24716 | 20.79 | GO:0032502 | 4265 | 3.59 |
| GO:0009295 | 5087 | 4.28 | GO:0005198 | 9151 | 7.70 | GO:0022610 | 1386 | 1.17 |
| GO:0031012 | 1555 | 1.31 | GO:0005215 | 2036 | 1.71 | GO:0032501 | 8520 | 7.17 |
| GO:0005576 | 43640 | 36.72 | GO:0016530 | 6280 | 5.28 | GO:0048518 | 2405 | 2.02 |
| GO:0030054 | 7902 | 6.65 | GO:0030545 | 10910 | 9.18 | GO:0040011 | 4423 | 3.72 |
| GO:0031974 | 25094 | 21.11 | GO:0003700 | 4277 | 3.60 | GO:0001906 | 31596 | 26.58 |
| | | | GO:0060089 | 1628 | 1.37 | GO:0048519 | 29664 | 24.96 |
| | | | GO:0016209 | 11287 | 9.50 | GO:0051704 | 10222 | 8.60 |
| | | | GO:0030234 | 14411 | 12.12 | GO:0022414 | 2495 | 2.10 |
| | | | | | | GO:0040007 | 2309 | 1.94 |
| | | | | | | GO:0002376 | 4514 | 3.80 |

In the CC domain, the number of selected protein sequences for GO: 0005576 ('extracellular region') is the largest, accounting for 36.72%, while the number for GO:0055044 ('symplast') is the smallest, only 0.99%. In the MF domain, the number of proteins belonging to GO:0016247 with function 'channel regulator activity' is the largest in non-annotated protein sequences, accounting for 20.79%. The number for GO: 0005085 ('guanyl-nucleotide exchange factor activity') is smallest, only 0.86%. From Ancestor Chart, the function 'antioxidant activity' (DO: 0016209) is a part of 'cellular response to stimulus' (GO: 0051716), which is closely related to the disease-resistance of creeping bentgrass. Its predicted number accounts for 9.50%. In the BP domain, GO: 0001906 ('Cell killing') has the largest number, accounting for 26.58%, followed by DO:0048519 ('negative regulation of biological process'), accounting for 24.96%. 3.8% of non-annotated protein sequences belongs to GO:0002376 ('immune system process').

## 3.4 Functional annotation of non-annotated sequences of creeping bentgrass by GO terms relevant to the disease-resistance response and signal transduction

In the above, the established prediction model was applied to select the protein sequences from non-annotated sequences of creeping bentgrass belonging to the GO terms chosen to establish the model. In this subsection, we focused on 13 GO terms with functions relevant to stimulus response and signal transduction related proteins (which was not used when establishing the prediction model). The model was trained by protein sequences of these GO terms, which were also collected from the UniPort database. The performance of the model is presented in Table 2. It is well known that the disease-resistance response and signal transduction process belong to BP domain. One can expect that the results are analogous to the results in the BP domain in Fig. 3(b). The SP values are considerable large, and spans from 96% up to 100%, which satisfies high false positive control rate required in the current work.

**Table 2**
The performance of prediction model and the number of the proteins for disease-resistance and signal transduction related GO terms. The GO ID, term, and number $N_{GO}$ of protein sequences in a GO term collected from the UniPort database are listed in the first to third columns, respectively. The measurements are listed in the fourth to eighth columns. The number of the proteins belonging to a GO term predicted from non-annotated protein sequences of creeping bentgrass is listed in last column.

| DO ID | Function | $N_{GO}$ | SE% | SP% | PR% | AC% | MCC | Num |
|---|---|---|---|---|---|---|---|---|
| GO:0009968 | negative regulation of signal transduction | 1730 | 42.3 | 97.3 | 84.4 | 83.0 | 0.52 | 148 |
| GO:0032102 | negative regulation of response to external stimulus | 542 | 32.7 | 99.4 | 94.9 | 82.0 | 0.49 | 11555 |
| GO:0044092 | negative regulation of molecular function | 1944 | 38.4 | 97.6 | 83.3 | 83.7 | 0.49 | 47437 |
| GO:0032101 | regulation of response to external stimulus | 1486 | 41.9 | 98.6 | 90.3 | 84.8 | 0.55 | 792 |
| GO:0002682 | regulation of immune system process | 2368 | 44.0 | 96.3 | 79.8 | 83.1 | 0.51 | 756 |
| GO:0009607 | response to biotic stimulus | 3184 | 35.2 | 98.1 | 86.6 | 81.8 | 0.48 | 24287 |
| GO:0006955 | immune response | 3214 | 34.7 | 97.7 | 83.1 | 82.4 | 0.46 | 1041 |
| GO:0009719 | response to endogenous stimulus | 2334 | 44.8 | 96.8 | 81.4 | 84.6 | 0.53 | 1035 |
| GO:0048585 | negative regulation of response to stimulus | 2341 | 35.9 | 97.4 | 83.7 | 80.9 | 0.46 | 402 |
| GO:0002764 | immune response-regulating signaling pathway | 611 | 47.7 | 96.9 | 84.7 | 84.0 | 0.55 | 1260 |
| GO:0044093 | positive regulation of molecular function | 2337 | 37.9 | 97.3 | 81.3 | 83.0 | 0.48 | 786 |
| GO:0051606 | detection of stimulus | 913 | 75.7 | 99.6 | 98.5 | 94.1 | 0.83 | 12 |
| GO:0080135 | regulation of cellular response to stress | 3819 | 68.4 | 96.4 | 85.7 | 89.8 | 0.70 | 3022 |

With the model trained by the data from UniPort, the protein sequences annotated from the non-annotated sequences of the creeping bentgrass are provided in the Additional file 4. In Table 2, we list the number of protein sequences with certain function. The number of protein sequences with function of 'negative regulation of molecular function' (GO:0044092) is the largest, accounting for 47,437. In Ancestor Chart, 'negative regulation of molecular function' is the biological regulation process. There are 24,287 annotated proteins with function of 'response to biological stimulus' (GO:0009607), 11,555 annotated proteins with function of 'negative regulation of response to external stimulus' (GO:0032102), and 3022 annotated proteins with function of 'regulation of cellular response to stress' (GO:0080135). These protein functions are closely related to the disease-resistance response of creeping bentgrass, and also reflect the positive response of creeping bentgrass to external stimulation after being infected by *Rhizoctonia solani* by inducing a large number of disease-resistance related proteins. Besides, there are 1260 annotation proteins with function of 'immune response regulating signaling pathway' (GO:0002764) and 148 annotation proteins with function of 'negative regulation of signal transduction proteins' (GO:0009968), both of which are closely related to signal transduction process. The annotation of the above protein functions is of the great significance for further explore the disease-resistance and signal transduction of creeping bentgrass.

## 4. Conclusions

To understand the ISR mechanism of turfgrass, the high-quality protein sequences were obtained from the creeping bentgrass seedlings infected by *Rhizoctonia solani* with BDO treatment to induce the ISR response. Amongst protein sequences obtained, some were functionally annotated according to the database alignment while the rest of the protein sequences were left non-annotated. A functional prediction model was established based on CNN with emphasizing high false positive control rate following Ref. [19], and applied to treat these left non-annotated proteins sequences to find the sequences relevant to the disease-resistance response and signal transduction which play important roles in ISR mechanism.

To establish the model, the data sets were collected from UniPort database in 29 GO terms in three domains. The sequences were encoded into a matrix, and convolution kernels were adopted to extract the information of sequences in different length levels. With the classification probability obtained, the protein sequences can be annotated. Compared the annotations with current model and these from UniPort database for the sequences in testing dataset, the prediction model was evaluated by five measurements, SE, SP, PR, AC, and MCC. A significant performance of the current model is the high SP values obtained for the data sets in 29 GO terms in three domains, which reflect the good false positive control rate. It guarantees the correctness of the annotation, though some sequences with certain function maybe omitted. With the established and trained model, the protein sequences belonging to 29 GO terms were selected from the non-annotated sequences of creeping bentgrass.

The established model was retrained by the data from UniPort in 13 GO terms relevant to the ISR mechanism. The non-annotated protein sequences of the creeping bentgrass were analyzed with retrained model. The protein sequences were annotated as different functions, which mainly involve 'response to biological stimulus', 'negative regulation of response to external stimulus', 'negative regulation of molecular function', 'regulation of cellular response to stress' and 'immune response regulating signaling'. These protein molecules play different roles in the disease-resistance process of creeping bentgrass. The results provide good candidates of the proteins with certain functions from all obtained protein sequences, which can be studied by molecular biology technology in further studies. With selected protein sequences, the waste of experimental data of protein sequence of creeping bentgrass can be avoided, and the experiment consumption of time and labor in further molecular biological studies can be saved. The current results are helpful to understand the ISR mechanism, and also provide reference for other sequence analysis of turfgrass disease-resistance research.

# Supplementary Information

**Additional file 1:** model's performance with GO terms in three domains.
**Additional file 2:** non-annotated amino acid sequences of creeping bentgrass.
**Additional file 3:** protein sequences selected by prediction model with the function of GO terms in three domains.
**Additional file 4:** protein sequences selected by prediction model with the function of GO terms relevant to the disease-resistance response and signal transduction.

# Declarations

**Ethics approval and consent to participate**

The experiments did not involve endangered or protected species. Experimental research and field studies on plants was carried out with permission of related institution, and complied with national or international guidelines and legislation.

**Consent for publication**

Not applicable.

**Availability of data and material**

Data generated or analyzed during this study were included in this published article and its supplementary information files. The sequences of creeping centgrass were uploaded on the National Center for Biotechnology Information (NCBI) Sequence Read Archive (SRA) under accession number SRR5658390. The sequences used in training were download from Uniport by GO terms directly.

**Competing interest**

The authors declare no conflict of interest.

**Funding**

This project is partially supported by the National Natural Science Foundation of China under Grant No. 11675228 and China postdoctoral Science Foundation under Grant No. 2015M572662XB.

**Author contributions**

Han-Yu Jiang and Jun He contributed to the conception of the study; Han-Yu Jiang performed the experiment to obtain the protein sequences of the creeping bentgrass; Jun He contributed significantly to establish the prediction model; Han-Yu Jiang and Jun He performed the data analyses and wrote the manuscript.

**Acknowledgements**

Not Applicable.

## Declaration of competing interest

The authors declare no conflict of interest.

## References

[1] D. Walters, A. Newton, G. Lyon, Induced resistance for plant defence. A sustainable approach to crop protection, Oxford: Blackwell Publishing (2007)

[2] C. X. Zhao, H. Y. Jiang, W. K. Dong, H. Chen, Y. X. Fanf, L. P. Xie, H. L. Ma. Effects of composite exogenous hormone application on induction of systemic resistance to Rhizoctonia solani in creeping bentgrass, Acta Prataculturae Sinica 27 (2018) 120-130.

[3] P.A.H.M. Bakker, R.V. Vanpeer, Suppression of soil-borne plant pathogens by fluorescent pseudomonads: Mechanisms and prospects. In biotic interactions and soil-borne diseases; A.B.R. Beemster, G.J.Bollen, Eds.; Elsevier Scientific: Amsterdam, the Netherlands (1991) 217–230.

[4] L.C. VanLoon, P.A.H.M. Bakker, Signaling in rhizobacteria-plant interactions. In root ecology, J. DeKroon, E.J.W. Visser, Eds. Springer: Berlin, Germany (2003) 287–330.

[5] M. Knoester, C.M.J. Pieterse, J.F. Bol, L.C. Van Loon, Systemic resistance in Arabidopsis induced by rhizobacteria requires ethylene-dependent signaling at the site of application, Mol. Plant-Microbe In. 12 (1999) 720-727.

[6] A.M. Cortes-Barco, T. Hsiang, P.H. Goodwin, Induced systemic resistance against three foliar diseases of agrostis stolonifera by (2R, 3R)-Butanediol or an isoparaffin mixture, Ann. of Applied Biol. Plant Pathol. 157 (2010) 179-189.

[7] H. Takahashi, T. Ishihara, S. Hase, A. Chiba, K. Nakaho, T. Arie, T. Teraoka, M. Iwata, T. Tugane, D. Shibata, S. Takenaka, Beta-cyanolanina synthase as a molecular marker for induced resistance by fungal glycoprotein elicitor and commercial plant activators, Phytopathology 96 (2006) 908-916.

[8] L. D. Bruyne, M. Höfte, D. D. Vleesschauwer. Connecting growth and defense: the emerging roles of brassinosteroids and gibberellins in plant innate immunity, Mol. Plant 7 (2014) 943–959.

[9] S. Kim, I.P. Ahn, Y.H. Lee, Analysis of genes expressed during rice - Magnapor the grisea interactions, Mol. Plant Microbe Interact. 14 (2001) 1340-1346.

[10] S.T. Kim, K.S. Cho, S. Yu, S.G. Kim, J.C. Hong, C.D. Han, D.W. Bac, M.H. Nam, K.Y. Kang, Proteomic analysis of differentially expressed proteins induced by rice blast fungus and clicitor in suspension-cultured rice cells, Proteomics 3 (2003) 2368-2378.

[11] I.S. Oh, A.R. Park, M.S. Bae, S.J. Kwon, Y.S. Kim, J.E. Lee, N.Y. Kang, S. Lee, H. Cheong, O.K. Park, Secretome analysis reveals an Arabidopsis lipase involved in defence against Alternaria brassicicola, The Plant Cell 17 (2005) 2832-2847.

[12] L.F. Thatcher，J.P. Anderson, K.B. Singh, Plant defence responses: what have we learnt from Arabidopsis? Funct. Plant Biol. 32 (2005) 1-19.

[13] H.Y. Jiang, J.L. Zhang, J.W. Yang, H.L. Ma, Transcript profiling and gene identification involved in the ethylene signal transduction pathways of creeping bentgrass (*Agrostis stolonifera*) during ISR response induced by butanediol, Molecules 13 (2018) 706.

[14] The Uniprot C. UniProt: the Universal Protein Knowledgebase, Nucleic Acids Res. (2017) 45(D1): D158-D69.

[15] M. Frasca, N. C. Bianchi, Multitask protein function prediction through task dissimilarity, Ieee Acm. T. Comput. Bi. 16 (2019) 1550-60.

[16] Y.X. Jiang，T.O. Ronnen，T. R. Wyatt Clark， B. Asma，R. Predrag, An expanded evaluation of protein function prediction methods shows an improvement in accuracy, Genome Biol. 17 (2016) 184.

[17] X. Pan, Y. Yang, C. Q. Xia, A.H. Mirza, H.B. Shen, Recent methodology progress of deep learning for RNA-protein interaction prediction, Wiley Interdiscip Rev. Rna 10 (2019) 1544.

[18] L. J. Colwell, Statistical and machine learning approaches to predicting protein-ligand interactions, Curr. Opin. Struct. Biol. 49 (2018) 123-8.

[19] J. Hong, Y. Luo, Y. Zhang, J.Ying, W. Xue, T. Xie, L. Tao, F. Zhu, Protein functional annotation of simultaneously improved stability, accuracy and false discovery rate achieved by a sequence-based deep learning. Brief. Bioinform. 21 (2019)1437-1447

[20] G.M.L.W Kroes, E. Sommers, Two in vitro assays to evaluate resistance in Linum usitatissimum to Fusarium wilt disease. Eur. J. Plant Pathol. 104 (1998) 561–568.

[21] Y. Kim, Convolutional neural networks for sentence classification, Proceedings of the 2014 Conference on Empirical Methods in Natural Language Processing (EMNLP) (2014) 1746–1751, arXiv:1408.5882

[22] Hamm CA, Wang CJ, Savic LJ, et al. Deep learning for liver tumor diagnosis part I: development of a

convolutional neural network classifier for multi-phasic MRI. Eur Radiol, 29 (2019) 3338–47.
[23] K. He, X. Zhang, S. Ren, and J. Sun, Delving deep into rectifiers: surpassing human-level performance on imageNet classification, Proceedings of the 2015 IEEE International Conference on Computer Vision (ICCV) (ICCV '15). IEEE Computer Society, USA (2015) 1026–1034, arXiv: 1502. 01852
[24] S.F. Altschul, T.L. Madden, Gapped BLAST and PSI-BLAST: A new generation of protein database search programs. Nucleic Acids Res. 25 (1997) 3389–3402.
[25] X. Mao, T. Cai, Automated genome annotation and pathway identification using the KEGG Orthology (KO) as a controlled vocabulary. Bioinformatics 21 (2005) 3787–3793.